\documentclass[11pt,a4paper,english]{article}
\usepackage[T1]{fontenc}
\usepackage[latin9]{inputenc}
\usepackage{amsthm}
\usepackage{amsmath}
\usepackage{amssymb}

\makeatletter

\providecommand{\tabularnewline}{\\}

\numberwithin{equation}{section}
\numberwithin{figure}{section}
\newcommand{\lyxaddress}[1]{
\par {\raggedright #1
\vspace{1.4em}
\noindent\par}
}
\newenvironment{lyxcode}
{\par\begin{list}{}{
\setlength{\rightmargin}{\leftmargin}
\setlength{\listparindent}{0pt}
\raggedright
\setlength{\itemsep}{0pt}
\setlength{\parsep}{0pt}
\normalfont\ttfamily}%
 \item[]}
{\end{list}}

\usepackage{babel}\usepackage{palatino}\usepackage{amsfonts}\makeatletter

\def\imod#1{\allowbreak\mkern10mu({\operator@font mod}\,\,#1)}

\makeatother

\makeatother

\usepackage{babel}

\begin{document}

\title{Affine Constellations Without Mutually Unbiased Counterparts}

\author{{\normalsize Stefan Weigert$^{\triangledown}$ and Thomas Durt$^{\vartriangle}$ }}

\maketitle

\lyxaddress{\begin{center}
$^{\triangledown}$Department of Mathematics, University of York,
UK-York YO10 5DD $^{\vartriangle}$IR-TONA, VUB, BE-1050 Brussels
\par\end{center}}
\begin{center}
s\texttt{low500@york.ac.uk,thomdurt@vub.ac.be}
\par\end{center}

\begin{abstract}
It has been conjectured that a complete set of mutually unbiased bases
in a space of dimension $d$ exists if and only if there is an affine
plane of order $d$. We introduce \emph{affine constellations} and
compare their existence properties with those of mutually unbiased
constellations. The observed discrepancies make a deeper relation
between the two existence problems unlikely. 
\end{abstract}
\global\long\def\kb#1#2{|#1\rangle\langle#2|}

\global\long\def\bk#1#2{\langle#1|#2\rangle}

\global\long\def\ket#1{|#1\rangle}

\global\long\def\bra#1{\langle#1|}

\global\long\def\c#1{\mathbb{C}^{#1}}

Two orthonormal bases $B$ and $B^{'}$ of a $d$-level quantum system
are \emph{mutually unbiased} (MU) if $|\bk b{b^{'}}|^{2}=1/d$ for
any two states $\ket b\in B$ and $\ket{b^{'}}\in B^{'}$. This means
that the probabilities for a transition of a quantum system prepared
in the state $\ket b\in B$ into a state $\ket{b^{'}}\in B^{'}$ are
independent of both the initital and the final state \cite{schwinger60}. 

It is known how to construct \emph{triples} of MU bases in $\c d$
for all values of $d\geq2$ \cite{grassl04}. The construction of
$(d+1)$-tuples of MU bases \cite{ivanovic81,wootters+89,bandyopadhyay+01}
can be based on Galois number fields or, alternatively, on fundamental
number-theoretical identities both of which, however, only exist if
the number $d$ is a prime or a power of a prime. Since the state
space $\c d$ of a $d$-level quantum system acccomodates at most
$(d+1)$ MU bases, the question of how many MU bases exist in spaces
of \emph{composite} dimension such as $d=6,10,12,\ldots$ arises naturally.
While it seems unlikely that composite dimensions support\emph{ complete}
sets of MU bases, no definite answer has yet been found. Spaces of
composite dimension appear to be {}``more generic'' (or {}``less
special'') than those of prime or prime power dimension---the known
constructions of complete sets of MU bases (see \cite{durt+10} for
a review) are based on mathematical structures existing for prime
(power) dimensions only. 

In dimension six, extensive numerical studies \cite{butterley+07,brierley+08}
support the view that certain subsets of $(d+1)$-tuples of MU bases,
known as \emph{MU constellations}, do not exist. Clearly, if a complete
set of seven MU bases were to exist in $\c 6$, then any MU constellation
obtained by removing some of these $d(d+1)$ vectors would exist as
well. Consider, for example, the MU constellation $\{5,5,3,1\}_{6\;}(\equiv\{5^{2},3,1\}_{6})$.
It consists of four sets of \emph{orthonormal} vectors in $\c 6$
containing $5,5,3,$ and $1$ elements, respectively, and the squared
modulus of the scalar product between vectors taken from \emph{different}
sets equals $1/6$. While this MU constellation of $14$ vectors has
been identified by numerical searches, other MU constellations with
the same number of vectors such as $\{5,4,3,2\}_{6}$ or $\{5,3^{3}\}_{6}$
have not been found, in spite of numerical efforts considerably larger
than those needed to identify $\{5^{2},3,1\}_{6}$. 

In an attempt to get a handle on the existence problem for complete
MU bases, it is natural to search for existence problems similar in
spirit. A promising candidate are \emph{finite affine planes }\cite{bennet95}:
these geometric structures consist of a finite number of\emph{ points}
which satisfy the following postulates: (i) any two points determine
a unique line; (ii) given a line and a further point, there is a unique
line through this point disjoint from the given line. Trivial realizations
of finite affine planes are excluded by the requirement that (iii)
there exist four points such that no three of them are located on
a single line. The \emph{order} $d$ of an affine plane is given by
the number of points on each line, and the entire plane can be \emph{foliated}%
\footnote{A foliation is also know as a \emph{striation}, or as a \emph{pencil}
in the mathematical literature.\emph{ }%
} into $d$ parallel (i.e. non-intersecting) lines in $(d+1)$ different
ways. 

Affine planes are readily constructed in terms of Galois fields if
their order $d$ is a prime number or a prime power \cite{wootters04},
in striking analogy to the known constructions of MU bases \cite{durt05}.
The Bruck-Ryser theorem \cite{bruck+49} shows that affine planes
of specific composite orders, $d=6,14,21,\ldots,$ do not exist, and
computer-aided combinatorics rule out the existence of an affine plane
of order $10$ \cite{lam+89}. 

The possibility of a link between the existence problems for MU bases
and affine planes has been voiced repeatedly \cite{zauner99,bengtsson+04},
with Wootters suggesting the explicit correspondence \cite{wootters04,wootters05}\emph{
}that\emph{ parallel lines of an affine plane should correspond to
operators projecting on orthogonal quantum states}. Consequently,
foliations are associated with orthonormal bases. Saniga, Planat,
and Rosu \cite{saniga+04} have elevated the link between MU bases
and affine planes to a conjecture: 
\begin{quotation}
\emph{{}``Non-existence of a projective plane of the given order
$d$ implies that there are less than $d+1$ mutually unbiased bases
(MUBs) in the corresponding $\mathcal{H}^{d}$, }and vice versa.''
\end{quotation}
A \emph{projective} plane of order $d$ turns into an affine plane
of the same order if one line is discarded%
\footnote{Conversely, an affine plane of order $d$ can be promoted to a projective
plane of the same order.%
}, thus covering the original claim. 

In this note, we will investigate the relation between MU bases and
affine planes in the light of MU constellations. Our observations
will suggest that the two existence problems actually exhibit \emph{less}
structural similarity (at least in dimension six) than one would hope
for on the basis of the SPR-conjecture.

Let us introduce the main concept needed for our argument, defined
in analogy to MU constellations: an \emph{affine constellation} $\langle x_{0},x_{1},\ldots,x_{d}\rangle_{d}$
of order $d$ consists of $(d+1)$ sets of $x_{b}\in\{0,1,\ldots,d-1\}$
lines with $d$ points each such that (a) any two lines within each
set do not intersect and (b) any two lines from different sets have
exactly one point in common. This notion is easily understood by example.
Given an affine plane of order $d=3$, there are four different ways
to arrange all nine points on three non-intersecting (parallel) lines.
Ignoring, say, seven of these twelve lines (one foliation, two lines
of the second foliation and one each of the remaining two), we create
the affine constellation $\langle2^{2},1\rangle_{3\;}(\equiv\langle2,2,1\rangle_{3})$.
It consists of three sets of $2,2,$ and $1$ lines, respectively,
such that the lines within each set indeed have (a) no point in common
while (b) lines belonging to different sets share exactly one point. 

Clearly, if an affine plane of order $d$ exists, all affine constellations
obtained by removing one line or more also exist. If, however, for
a given value of $d$, some affine constellation is found \emph{not}
to exist, then an affine plane of order $d$ cannot exist. Assuming
that the SPR-conjecture captures a fundamental mathematical relationship
between MU bases and affine planes, one would expect the properties
of affine constellations to closely parallel those of MU constellations.

In prime power dimensions, parallel lines of an affine plane can be
associated successfully with operators projecting on orthogonal quantum
states. We now turn to the question whether this correspondence continues
to hold for affine constellations, especially in composite dimensions
such as $d=6$. 

Let us begin with a few encouraging observations. A first important
property of orthonormal bases of $\c d$ does have an analogue for
affine planes: there is only one way to complete $(d-1)$ orthonormal
vectors of $\c d$ into a basis%
\footnote{This is the reason for listing only $(d-1)$ states in a MU constellation:
$\{5^{3}\}_{6}$, for example, denotes three MU bases in dimension
six, containing six vectors each, although only $15$ vectors are
exhibited. %
} just as there is a unique line parallel to $(d-1)$ non-intersecting
lines of an affine plane%
\footnote{Thus, we will also limit the number of lines necessary to specify
an affine constellation to $(d-1)$, so that $\langle2^{4}\rangle_{3}$
denotes the affine plane of order three.%
} of order $d$. Second, this last line is easily seen to intersect
only once with any other line which has only one point in common with
each of the $(d-1)$ given lines. In the context of MU constellations,
this means that the $d^{th}$ line is automatically MU to all states
which are MU to the original $(d-1)$ states. Finally, given $d$
different foliations of $d^{2}$ points by $d$ lines, a $(d+1)^{st}$
foliation must exist (\cite{weiner09}, stated without proof%
\footnote{Given the affine constellation $\langle(d-1)^{d}\rangle_{d}$, consider
those $d$ lines of the $d$ foliations of the $d^{2}$ points that
pass through an arbitrary point $P$. There remain $d-1=d^{2}-(d(d-1)+1)$
points that do not belong to those lines. The union of these $d-1$
points with the point $P$ defines a line $L_{P}$. Repeating this
construction with a different point $P'$ we obtain a line $L_{P'}$
which either coincides with $L_{P}$ (if $P'$ belongs to $L_{P}$)
or is parallel to $L_{P}$ (if $P'$ is not on $L_{P}$). Ultimately,
we end up with a new foliation consisting of $d$ distinct, parallel
lines each of which, by construction, intersects any line of $\langle(d-1)^{d}\rangle_{d}$
in only one point. Adding this foliation to the initial affine constellation,
you obtain an affine plane $\langle(d-1)^{d+1}\rangle_{d}$ of dimension
$d$ since the postulates (i) to (iii) are satisfied.%
}) promoting the affine constellation $\langle(d-1)^{d}\rangle_{d}$
to the affine plane $\langle(d-1)^{d+1}\rangle_{d}$. An analogous
property holds for MU constellations: given $d$ MU bases, a \emph{complete}
set of $(d+1)$ MU bases $\{(d-1)^{d+1}\}_{d}$ can be constructed
\cite{weiner09}. Here the similarities end.

We now come to the main point of this note:
\begin{quote}
\emph{MU constellations and affine constellations do not match in
dimension six.} 
\end{quote}
To see this, let us consider $36$ points which are known not to support
an affine plane of order six---the maximal number of foliations is
three, not seven \cite{tarry00}. The largest possible affine constellation
(containing three foliations) is given by $\langle5^{3},4\rangle_{6}$
since adding a fifth line to the last four would imply the existence of \emph{four}
foliations, a contradiction. This constellation exists: two (standard) foliations of $\langle5^{3},4\rangle_{6}$
are given by the six horizontal and six vertical lines, and Table
1 makes the remaining ten lines explicit using the notation of a \emph{Graeco-Latin
square }for a pair of\emph{ mutually orthogonal Latin squares}, or
MOLS \cite{bennet95} (see the caption for details). 

\begin{table}
\begin{lyxcode}
\begin{centering}
\begin{tabular}{|c|c|c|c|c|c|}
\hline 
$54$ & $2\,\cdot$ & $3\,\cdot$ & $63$ & $11$ & $42$\tabularnewline
\hline 
$1\,\cdot$ & $53$ & $64$ & $4\,\cdot$ & $22$ & $31$\tabularnewline
\hline 
$2\,\cdot$ & $62$ & $51$ & $3\,\cdot$ & $44$ & $13$\tabularnewline
\hline 
$61$ & $1\,\cdot$ & $4\,\cdot$ & $52$ & $33$ & $24$\tabularnewline
\hline 
$32$ & $41$ & $23$ & $14$ & $5\,\cdot$ & $6\,\cdot$\tabularnewline
\hline 
$43$ & $34$ & $12$ & $21$ & $6\,\cdot$ & $5\,\cdot$\tabularnewline
\hline
\end{tabular}
\par\end{centering}
\end{lyxcode}
\caption{This (incomplete) Graeco-Latin square represents one foliation and
four additional lines of $\langle5^{3},4\rangle_{6}$, the maximal
affine constellation of order $6$. The first integer in each square
indicates one of the six lines of the (non-standard) foliation to
which the corresponding point belongs. These integers are different
in each row and column ensuring that each line has only one point
in common with the standard foliations consisting of horizontal and
vertical lines, respectively. Four more lines are defined by the second
integers which, again, they do not repeat within any line or column.
Finally, no two squares contain the same two-digit number to ensure
that each of the four lines intersects those of the third foliation
in a single point only.}

\end{table}

Now, a valid association of lines in an affine constellation with
projection operators acting in $\c 6$ would suggest the existence
of the MU constellation $\{5^{3},4\}_{6}$. However, as mentioned
before, there is strong evidence for the \emph{non-existence} of the
MU constellations $\{5,4,3,2\}$ and $\{5^{3},3\}_{6}$ which imply
the non-existence of $\{5^{3},4\}_{6}$ -- and, \emph{a fortiori},
of any MU constellation {}``in between'' since MU constellations
form a \emph{lattice} \cite{brierley+08}. Thus, while Table 1 explicitly
exhibits the affine constellation $\langle5^{3},4\rangle_{6}$, its
MU counterpart is unlikely to exist. This is our main result, and
it casts doubt on the (attractive) idea of a deeper structural relation
between affine constellations and MU constellations.

Let us point out further mismatches in dimension six: 
\begin{itemize}
\item The MU constellation $\{5^{3}\}_{6}$ consisting of three MU bases related
to the Heisenberg-Weyl group cannot be extended by a single MU vector
\cite{grassl04}. Thus, it is impossible to associate any of the four
lines in the affine constellation $\langle5^{3},4\rangle_{6}$ with
such a vector once the three foliations have been mapped to three
Heisenberg-Weyl type MU bases (cf. \cite{paterek+09} for a similar
argument in dimension $10$). 
\item The columns of the unit matrix and the\emph{ special} (or Tao's) matrix
define two MU bases which have been shown to be unextendible by two
orthonormal MU vectors, that is, to the MU constellation $\{5^{2},2\}_{6}$
\cite{brierley+09-1}. In contrast, nothing prevents us from extending
any two foliations of $36$ points by two lines to obtain the affine
constellation $\langle5^{2},2\rangle_{6}$.
\item The unit matrix and the matrices of the Fourier family \cite{Hadamardsonline}
provide a two-parameter set of inequivalent MU constellations $\{5^{2}\}_{6}$.
However, a discrete set of $36$ points cannot support \emph{continuous}
families of affine constellations $\langle5^{2}\rangle_{6}$. 
\end{itemize}
Similar disparities also arise for constellations in prime and prime
powers. Suffice it to recall that for $d=4$ the set of triples of
MU bases depends on three continuous parameters \cite{brierley+09}
and again, there is no room for continuous families of affine constellations.

Strictly speaking, these mismatches between affine and MU constellations
do not directly affect the SPR-conjecture since it only relates \emph{complete}
sets of MU bases to finite projective (and hence affine) planes, i.e
\emph{maximal} affine constellations. Nevertheless, the conjecture
is mathematically considerably less attractive if its natural extension
to constellations does not hold. We feel that the discrepancies just
described indicate the absence of a deeper structural similarity between
the two existence problems.%
\footnote{The Bruck-Ryser theorem, establishing the non-existence of finite
affine planes in specific composite dimensions, is a case in point:
closer inspection of the proof shows that it relies on entries of
certain matrices to take values of $0$ and $1$ only while the elements
of Hadamard matrices, their equivalent in the setting of MU bases,
are complex numbers of modulus one.%
}

These observations clearly leave room for alternative links between
af$\-$fine and MU constellations. It would be interesting, for example,
to establish a weaker correspondence between \emph{foliations} of
$d^{2}$ points and MU bases in $\c d$ for composite dimensions $d=\Pi_{n}p_{n}^{k_{n}}$
(with prime numbers $p_{n}$, positive integers $k_{n}$, and increasing
factors $p_{n}^{k_{n}}$). Any such correspondence would need to accomodate
a number of known facts. There exist $(p_{1}^{k_{1}}-1)$ MOLS,\emph{
}\cite{macneish21,bennet95}, giving rise to $(p_{1}^{k_{1}}+1)$
foliations from which one may construct $(p_{1}^{k_{1}}+1)$ MU bases
\cite{paterek+09-2}. In infinitely many \emph{square} dimensions
\emph{more} MU bases can be found \cite{wocjan+05}: for example,
four MOLS in dimension $d=2^{2}\cdot13^{2}$ translate into \emph{six}
MU bases. Interestingly, a particular construction of MU bases via
MOLS which works for prime power dimensions \emph{fails} in composite
dimensions such as $d=10$ \cite{paterek+09}. Furthermore, our observations
suggest that the correspondence might not be one-to-one in some cases.
The situation is complicated even further if one considers sets of
\emph{triples} of MU bases in low dimensions: for $d= 3$, there is only
\emph{one} triple of MU bases; for $d=4$, there is a \emph{three-parameter
family} of MU triples; and for $d=5$, there is a \emph{pair} of MU triples
\cite{brierley+09-1}.

In summary, we feel that the observed structural discrepancies result
from the fact that state space of affine constellations is \emph{finite}
while Hilbert space, the habitat of MU bases, has room for families
of states depending on \emph{continuous parameters}. Simply enumerating
all possible candidates of lines allows one to confirm the non-existence
of an affine plane \emph{par épuisement} (which, in fact, lead to
the first proof that there is no affine plane in dimension six \cite{tarry00}).
Similarly, the computer-aided exhaustive enumeration of a \emph{finite}
number of cases plays an important role in disproving the existence
of an affine plane of order $10$. 

It is true that promising approaches to prove the non-existence of
complete sets of MU bases reduce the problem to an \emph{algorithmic}
one such that only a finite number of cases need to be checked \cite{Jaming+09,brierley+10}.
However, this does not take away anything from the fundamental difference
between the state spaces of the two problems which, in our view, speaks
against the SPR-conjecture capturing a deep mathematical truth.

\subsection*{Acknowledgements}

We would like to thank H. Zainuddin for his hospitality during EQuaLS
3, where this work was initiated, and I. Bengtsson, S. Brierley, and
T. Sudbery for constructive comments on early drafts.

\end{document}